\documentclass[final,1p,times]{elsarticle} 
\usepackage{graphicx} 
\usepackage{amssymb} 
\usepackage{amsthm} 
\usepackage{lineno} 

\usepackage{hyperref}

\def\mean#1{\ensuremath{\left<#1\right>}}

\newcommand{\lambdas}{\ensuremath{\lambda^{*}}}
\newcommand{\lamfrac}{\ensuremath{\lambda^{*}(\mT)/\lambda_{max}^{*}}}
\newcommand{\mT}{\ensuremath{m_\mathrm{T}}}

\newcommand{\etap}{\ensuremath{\eta^\prime}}
\newcommand{\metap}{\ensuremath{m_{\etap}}}
\newcommand{\meps}{\ensuremath{m_{\etap}^{*}}}
\newcommand{\Tfo}{\ensuremath{T_{FO}}}

\newcommand{\Tcond}{\ensuremath{T_{cond}}}
\newcommand{\Binv}{\ensuremath{B^{-1}}}
\newcommand{\uT}{\ensuremath{\mean{u_\mathrm{T}}}}

\journal{Nuclear Physics A} 
\begin{document} 

\begin{frontmatter} 


\title{Significant in-medium $\etap$ mass reduction in $\sqrt{s_{NN}}=200$ GeV Au+Au collisions\footnote{Dedicated to Mikl\'os Gyulassy in celebration of his 60th birthday. Supported by OTKA NK 73143 and T049466.}}

\author{R. V\'ertesi$^a$, T.~Cs\"org\H{o}$^{a,b}$ and J.~Sziklai$^a$}

\address{$^a$ MTA KFKI RMKI, Budapest, H - 1525, Hungary \\
         $^b$ Dept. Physics, Harvard University, 17 Oxford St, Cambridge, MA 02138, USA}

\begin{abstract} 
PHENIX and STAR data on the intercept parameter of 
the two-pion Bose-Einstein correlation functions in $\sqrt{s_{NN}}= 200$ GeV Au+Au collisions
were analysed in terms of various models of hadronic abundances.
To describe these data, an in-medium $\eta^\prime$ mass decrease of at least 200 MeV was needed in these models.
\end{abstract} 

\end{frontmatter} 



\begin{figure}[tbp]
\begin{center}
\includegraphics[height=45mm]{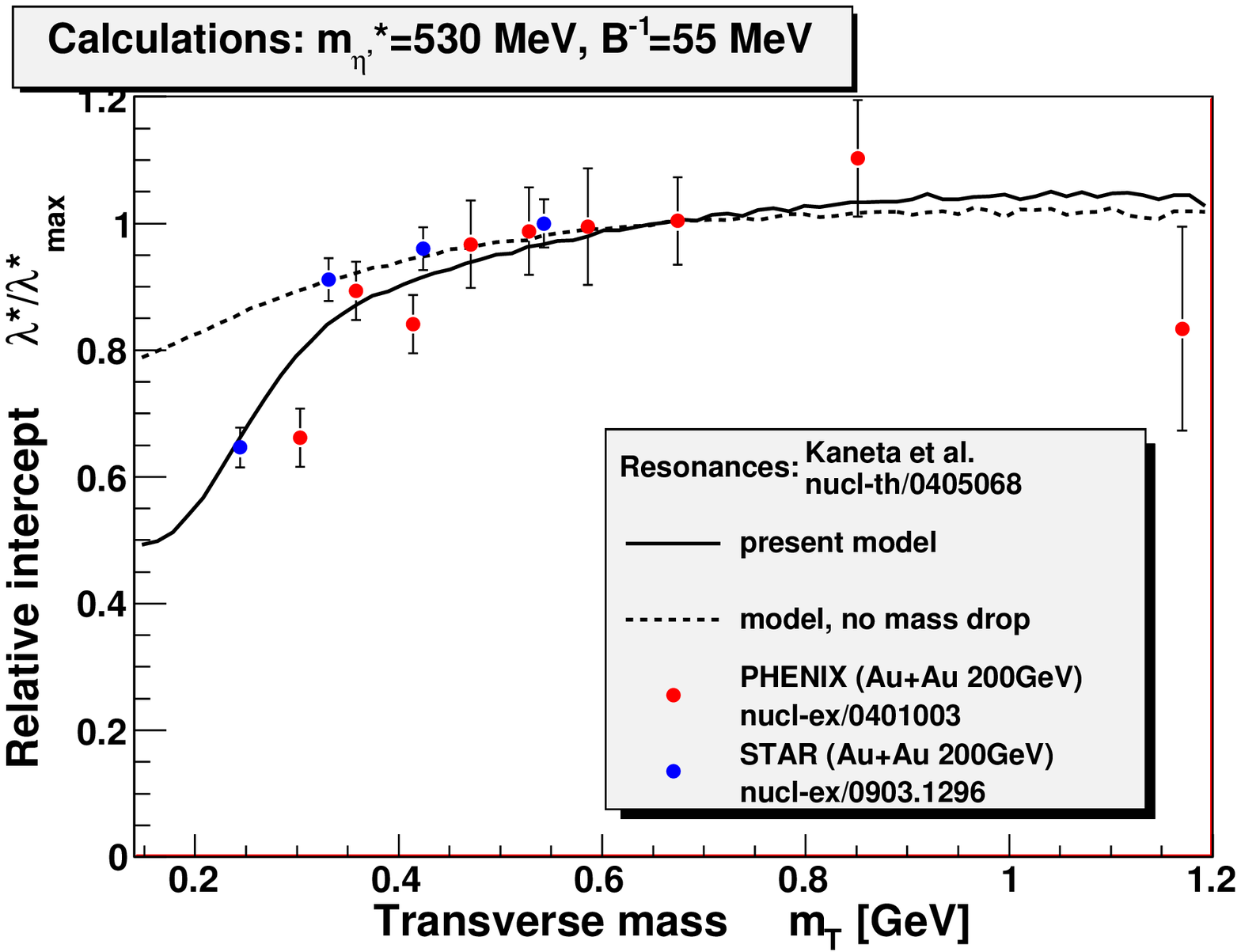}
\includegraphics[height=45mm]{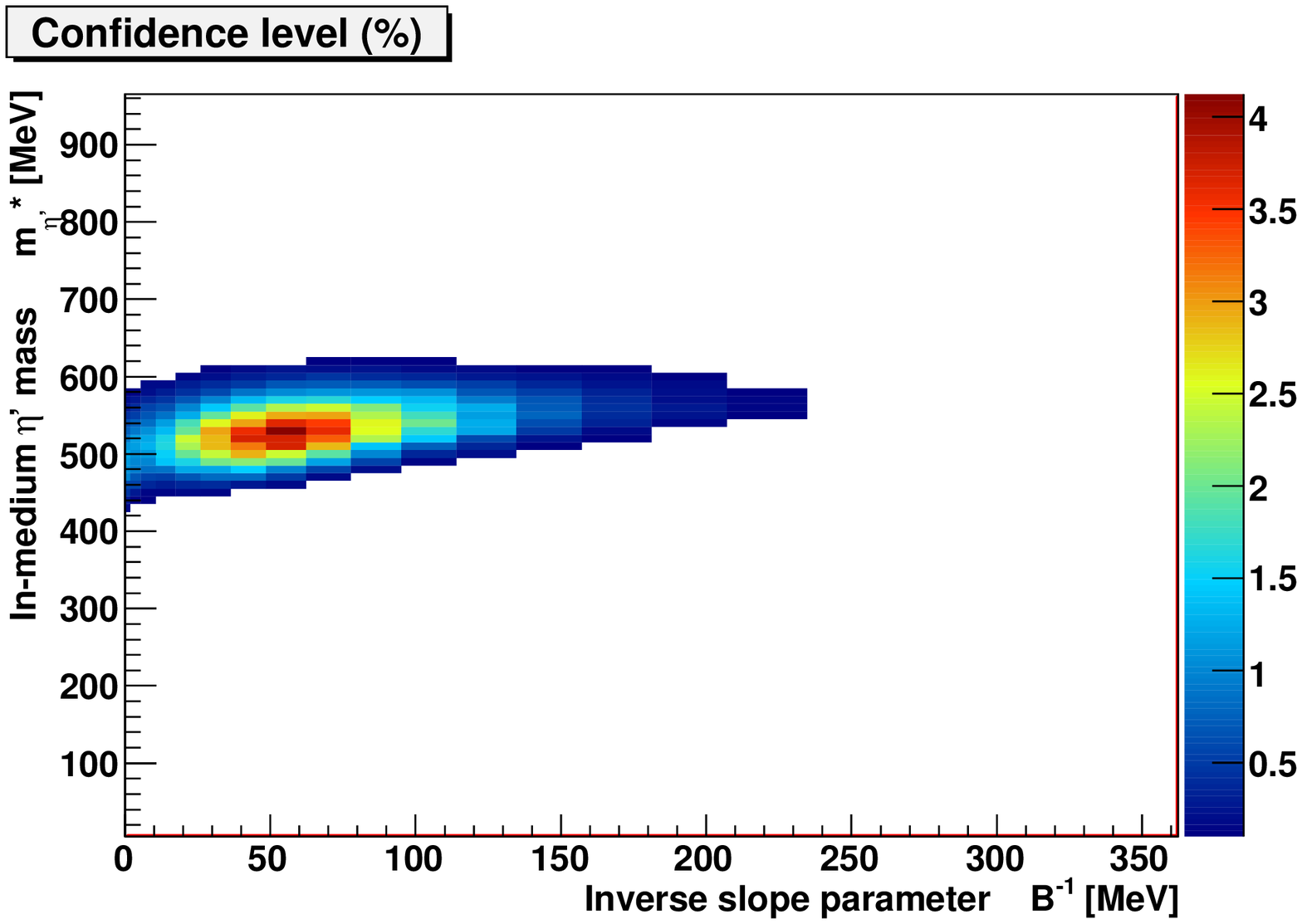}
\end{center}
\vspace{-0.5cm}
\caption{
{\it (Left)} 
Monte Carlo simulations of  
\lamfrac\ compared with PHENIX and STAR data. 
{\it (Right)} 
Confidence level distibution of these simulations, 
at various values of the in-medium $\eta^{\prime}$  mass
and slope parameter of the $\eta^\prime$ condensate 
$\Binv{}$, for $\alpha=0$, $\Tfo=\Tcond=177$ MeV and $\uT=0.48$ \cite{vance}.
Resonance ratios of ref.~\cite{kaneta} were utilized in both panels.
}
\label{fig:kaneta}
\end{figure}
\begin{figure}[tbp]
\begin{center}
\includegraphics[height=45mm]{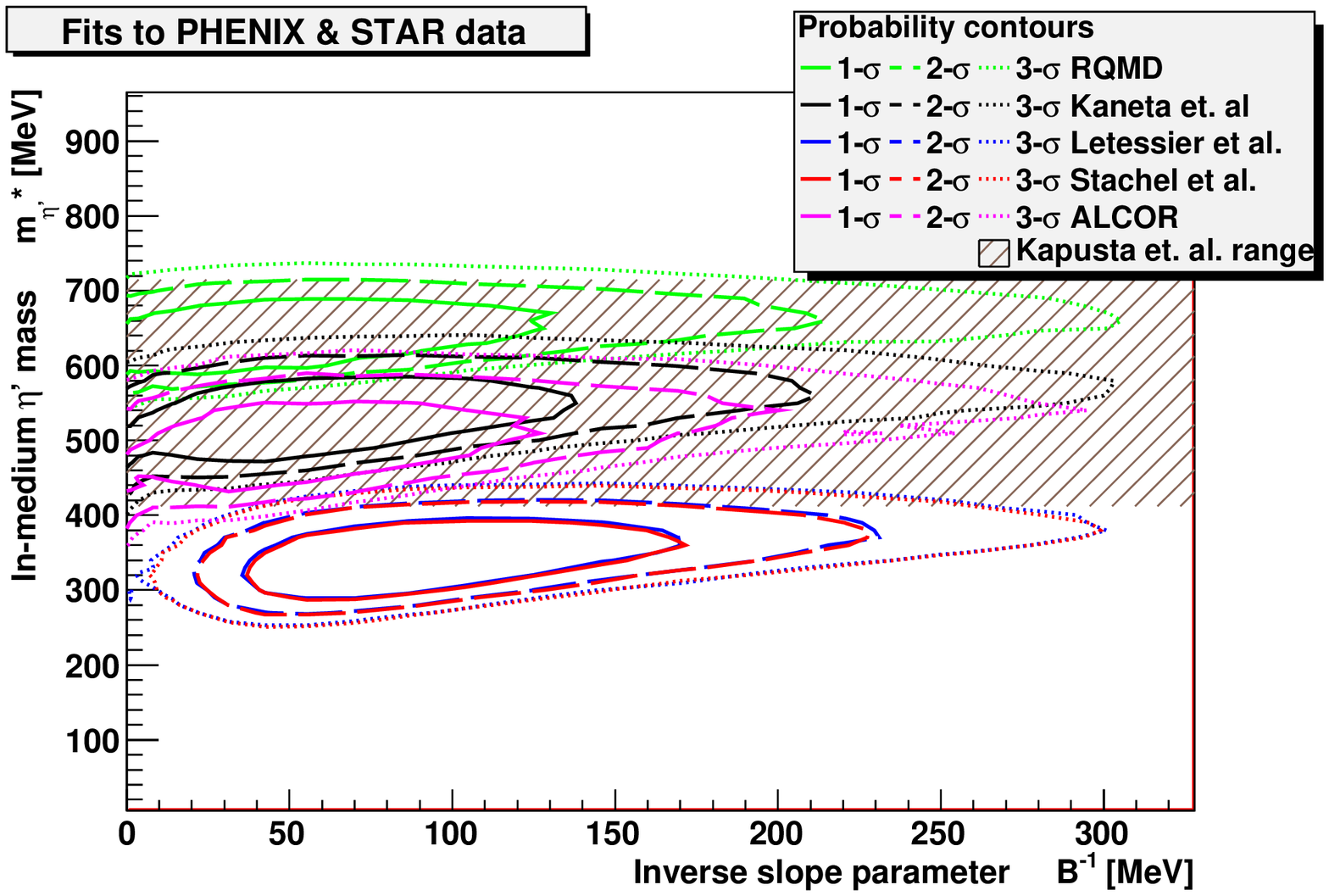}
\includegraphics[height=45mm]{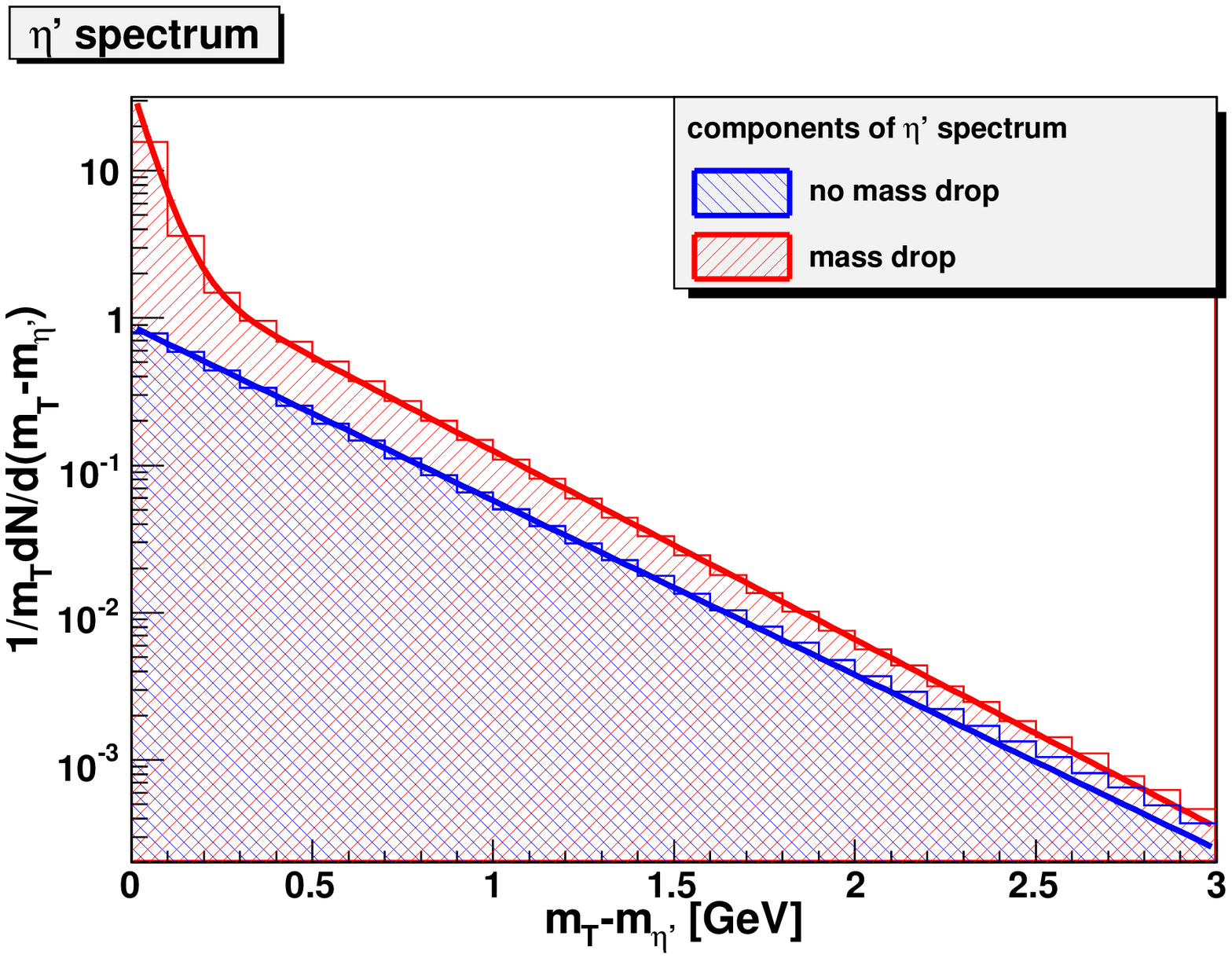}
\end{center}
\vspace{-0.5cm}
\caption{
{\it (Left)}
Standard deviation contours on the (\Binv{}, \meps) plain, obtained from 
\lamfrac\ of MC simulations based on particle abundances of~\cite{alcor,rqmd,kaneta,rafelski,stachel}, each fitted
successfully to the PHENIX and STAR combined dataset, while
fits based on ref.~\cite{fritiof} were statistically not acceptable
and are not shown.
The dashed band indicates the theoretically predicted range~\cite{kapusta}.
{\it (Right)} Reconstructed $\mT$ spectrum of the $\eta^\prime$ mesons. 
Lower part indicates the scenario without in-medium $\eta^\prime$ mass reduction, 
upper part the enhancement required to describe
the dip in the low $\mT$ region of $\lambdas$.
}
\label{fig:map_cl}
\vspace{-0.5cm}
\end{figure}

In high energy heavy ion collisions, a hot and dense medium is created, where the $\mathrm{U_A}(1)$ 
or chiral symmetry may temporarily be restored \cite{kunihiro,kapusta,huang}. 
As a consequence, the mass of the ``prodigal" $\etap(958)$ mesons~\cite{kapusta} may be reduced
to its quark model value and the abundancy of these $\etap$ mesons at low $p_T$ may
be enhanced by more than a factor of 10. 
The transverse mass ($m_T$) dependence of the intercept parameter 
$\lambdas$ of the charged pion Bose Einstein Correlations 
provides an  observable which is sensitive to such enhanced $\etap$ abundancy \cite{vance}. 
We have analysed PHENIX and STAR data on the relative strength of $\lambdas(\mT)/\lambdas_{max}$ 
\cite{phnxpub,starpub} using extensive Monte Carlo simulations based 
on various models (ALCOR, FRITIOF, RQMD, and thermal models by Kaneta, Rafelski, Stachel and collaborators)
for hadronic abundances \cite{alcor,fritiof,rqmd,kaneta,rafelski,stachel}.
Resonance decays were performed with JETSET 7.4~\cite{Sjostrand:1995iq}.
Our simulations improved those of \cite{vance,phnxpre}: 
The number of in-medium $\etap$ mesons was calculated with an improved Hagedorn formula,  which included a prefactor with an expansion dynamics dependent exponent $\alpha$: 
\begin{equation}\label{eq:prietamtdist}
f=\left(\frac{\meps}{\metap}\right)^\alpha e^{- \frac{\metap-\meps}{\Tcond}}.
\end{equation}
A slope parameter, $B^{-1}$ was introduced too, to describe the transverse mass spectra of the $\eta^\prime$ mesons produced when the condensated in-medium $\eta^\prime$-s come on-shell.
Systematic studies were carried out for various reasonable values of $\alpha$ and other model parameters like the $\etap$ freezeout temperature $\Tcond$.
These simulations with sufficiently large in-medium $\eta^\prime$ mass reduction 
described both PHENIX and STAR data (Fig.~\ref{fig:kaneta}). The best values for the in-medium mass
of $\eta^\prime$ mesons were in the theoretically predicted range~\cite{kapusta}, or slightly below it (Fig.~\ref{fig:map_cl} left panel).
The best parameter regions for the considered models are shown in the left panel
of Fig.~\ref{fig:map_cl}, while 
the low transverse momentum enhancement in the $\eta^\prime$ spectrum is shown in the right panel.

At the 99.9 \% confidence level~\cite{vertesi-web}, 
at least 200 MeV in-medium decrease of the mass of the $\eta^\prime(958)$ meson was needed 
in  the considered model class to describe both
STAR and PHENIX data on $\lambdas(\mT)/\lambda_{max}$ 
of $\sqrt{s_{NN}} = 200$ GeV Au+Au collisions.
\vspace{0.3cm}

\begin{thebibliography}{00} 
   \bibitem{kunihiro} T. Kunihiro, Phys. Lett. {\bf B219} 363 (1989); ibid. {\bf B245} 687(E) (1990).
\bibitem{kapusta} J. Kapusta, D. Kharzeev and L. McLerran, Phys. Rev. {\bf D53} 5028 (1996).
\bibitem{huang}   Z. Huang and X.-N. Wang, Phys. Rev. {\bf D53} 5034 (1996).
\bibitem{vance}   S. Vance, T. Cs\"org\H{o} and D. Kharzeev, Phys. Rev. Lett. {\bf 81} 2205 (1998).
\bibitem{phnxpub} S. Adler {\it et. al} [PHENIX collaboration], Phys. Rev. Lett. {\bf 93} 152302 (2004).
\bibitem{starpub} J. Adams {\it et al.} [STAR collaboration], Phys. Rev. {\bf C71} 044906 (2005).
\bibitem{alcor}   T.~S.~Bir\'o, P.~L\'evai and J.~Zim\'anyi, Phys.\ Lett.\  B {\bf 347}, 6 (1995).
\bibitem{fritiof} B. Anderson et al., Nucl. Phys. B {\bf 281} (1987) 289.
\bibitem{rqmd} J. P. Sullivan et al., Phys. Rev. Lett. {\bf 70} (1993) 3000.
\bibitem{kaneta}  M. Kaneta and N. Xu, nucl-th/0405068.
\bibitem{rafelski}J.~Letessier and J.~Rafelski, Eur.\ Phys.\ J.\  A {\bf 35}, 221 (2008).
\bibitem{stachel} S.~A.~Bass {\it et al.}, Nucl.\ Phys.\  A {\bf 661}, 205 (1999).
\bibitem{Sjostrand:1995iq} T.~Sj\"ostrand, Comp.\ Phys.\ Commun. {\bf 82} (1994) 74.
\bibitem{phnxpre} M. Csan\'ad for PHENIX Collaboration, Nucl. Phys. {\bf A774} 611-614 (2006).
\bibitem{vertesi-web} \url{http://www.rmki.kfki.hu/~vertesi/etap_bec/}.

\end{thebibliography}
\end{document}